\documentclass{article}

\usepackage{arxiv}

\usepackage[utf8]{inputenc} 
\usepackage[T1]{fontenc}    
\usepackage[hidelinks]{hyperref}       
\usepackage{url}            
\usepackage{booktabs}       
\usepackage{amsfonts}       
\usepackage{nicefrac}       
\usepackage{microtype}      
\usepackage{cleveref}       
\usepackage{lipsum}         
\usepackage{graphicx}
\usepackage{natbib}
\usepackage{doi}

\title{Ethical conundrums: Hacked data in the study of far-right violent extremism}

\date{7 November 2025}

\newif\ifuniqueAffiliation
\uniqueAffiliationtrue

\ifuniqueAffiliation 
\author{ \href{https://orcid.org/0000-0002-0643-1405}{\includegraphics[scale=0.06]{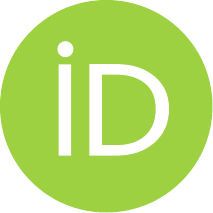}\hspace{1mm}Lise Waldek} \\
	Institute for Sustainable Industries and Livable Cities\\ Victoria University\\
    Melbourne VIC 3000 \\
	\texttt{lise.waldek@live.vu.edu.au} \\
	\And
	\href{https://orcid.org/0000-0003-4932-7912}{\includegraphics[scale=0.06]{orcid.pdf}\hspace{1mm}Brian Ballsun-Stanton} \\
	Faculty of Arts\\
	Macquarie University\\
	North Ryde NSW 2109\\
	\\
    \And
    \href{https://orcid.org/0000-0001-8002-0551}{\includegraphics[scale=0.06]{orcid.pdf}\hspace{1mm}Muhammad Iqbal}\\
	Institute for Sustainable Industries and Livable Cities\\ Victoria University\\
    Melbourne VIC 3000 \\
    \And
    \href{https://orcid.org/0000-0002-5024-7541}{\includegraphics[scale=0.06]{orcid.pdf}\hspace{1mm}David Kernot}\\
	Institute for Sustainable Industries and Livable Cities\\ Victoria University\\
    Melbourne VIC 3000 \\
    \And
    \href{https://orcid.org/0000-0002-4789-3489}{\includegraphics[scale=0.06]{orcid.pdf}\hspace{1mm}Debra Smith}\\
	Institute for Sustainable Industries and Livable Cities\\ Victoria University\\
    Melbourne VIC 3000 \\
}

\else

\fi

\renewcommand{\shorttitle}{Ethics of Hacked Data}

\hypersetup{
pdftitle={Ethical conundrums: Hacked data in the study of far-right violent extremism},
pdfsubject={cs.CY},
pdfauthor={Lise Waldek, Brian Ballsun-Stanton, Muhammad Iqbal, David Kernot, Debra Smith},
pdfkeywords={Hacked Data; Ethics; Violent Extremism; Privacy},
}

\begin{document}
\maketitle

\begin{abstract}
Ethical conduct in digital research is full of grey areas. Disciplinary,
institutional and individual norms and conventions developed to support
research are challenged, often leaving scholars with a sense of unease
or lack of clarity. The growing availability of hacked data is one area.
Discussions and debates around the use of these datasets in research are
extremely limited. Reviews of the history, culture, or morality of the
act of hacking are topics that have attracted some scholarly attention.
However, how to undertake research with this data is less examined and
provides an opportunity for the generation of reflexive ethical
practice. This article presents a case-study outlining the ethical
debates that arose when considering the use of hacked data to examine
online far-right violent extremism. It argues that under certain
circumstances, researchers can do ethical research with hacked data.
However, to do so we must proactively and continually engage deeply with
ethical quandaries and dilemmas.
\end{abstract}


\section{INTRODUCTION}

The release of public interest hacked datasets are increasing. There are
many ways to describe data published online in the absence of consent --
hacked, stolen, breached, leaked. This article explores hacked data --
``data that has been obtained in an unauthorised manner through illicit
access to a computer or computer network'' (Ienca and Vayena, 2021a).
Often the data is released online where it can be accessed by any
individual. Ethical guidance on the appropriateness of utilising these
data for research purposes is extremely slim, with the majority of
discussions simply alluding to the complexity of the ethical challenges
the data raises (Tiidenberg, 2017). This article presents a case study
where ethics approval for the use of hacked data was achieved for
purposes of studying the behaviours of online far-right violent
extremists\footnote{Ethics approval was granted by Victoria University,
  Australian reference -- HRE23-177}. This paper adopts a case study
approach to explore procedural ethics in the context of hacked online
data. While the use of case studies has been at times maligned in the
social sciences, others have robustly argued for the utility of this
approach when researching the complexities and contradictions of real
life (Flyvbjerg, 2006; White and Cooper, 2022). The use of a single case
study drawn from an Australian University provides us with the
opportunity to deeply engage with the theoretical ethical framework
proposed by Ienca and Vayena as a mechanism for the robust debates
required around the balancing of risks and benefits when seeking to use
hacked data. The paper argues that under certain circumstances
researchers can design ethical public interest research programs that
involve the use of hacked data. We suggest the six criteria provided by
Ienca and Vayena should be incorporated into future iterations of the
AoIR ethical guidelines as a starting point for researchers seeking to
engage with hacked data (Ienca and Vayena, 2021a). Making this argument
does not negate the deep ethical quandaries and obligations researchers
must grapple with.

Kate Crawford and danah boyd have championed the need for robust peer
debate on tricky ethical issues arising from the specificities of big
data research, noting how these discussions are a critical feature of
research (Zook et al., 2017). Likewise, the Association of Internet
Research (AoIR) ethical guidelines note how these discussions are likely
to ``evoke more than one ethically defensible response to a specific
dilemma or problem. \emph{Ambiguity, uncertainty, and disagreement are
inevitable.}'' (Internet Research Ethics 1.0 p. 3, emphasis in the
original). Awareness of the need for these debates within the
sub-discipline of online violent extremism and terrorism are also
growing. Maura Conway argues against prescriptive ethical guidance given
the inherent dynamism of the digital environment but notes this should
not prohibit robust and transparent debate. It is within the context of
these debates that we contribute to the conversation by providing an
example of the challenges we faced during our own ethics journey in
relation to a project examining far-right extremist online behaviours
and activities. In this sense we are responding to Maura Conway's call
for researchers to ``commit our ethical decisions and decision-making
processes to writing more often and fully\ldots so that a store of
usable knowledge is built up over time that is then useable by us in our
own and other's decision making including that of Research Ethics
Committees (REC)/Institutional Review Board's (IRB)'' (Conway, 2021:
377).

The paper begins by considering general risks often associated with the
design and implementation of ethical research involving social media and
online data. It then explores how these and other risks manifest when
considering the potential ethical use of hacked data. Drawing on a case
study methodology, the paper then explores how the theoretical ethical
framework proposed by Ienca and Vayena as a starting point for
researchers considering the use of hacked data, have successfully been
applied to a real-world research project. We are mindful that
researchers may seek actionable take-aways from this paper when
preparing their own ethics application to a potentially sceptical ethics
board. Therefore, we have structured our case study against the
high-level topic areas identified by Ienca and Vayena that guided our
own reflections and engagement with the ethics process. In addition, we
begin the consideration of these topic areas with questions that we
developed to guide our thinking on ethical issue. These could be applied
or adapted by researchers in the context of their own cultural,
geographical and theoretical positions as they develop their ethics
application for the use of hacked data. These additions and the
presentation of a real-world case study provide a novel and useful
structured way to assist in the rigorous debates and discussions
underpinning ethical decision making when weighing up the balancing of
risks and benefits in the use of hacked data for research purposes.

\section{GENERAL RISKS WITH SOCIAL MEDIA DATA}

Markham and colleagues described how the affordances of digital
technologies result in a ``transmogrification of the person as a
coherent representational entity into a constellation of data points
abstracted from social context and lived experience'' (Markham et al.,
2018: 1--2). The digital environment has problematised -- conceptually
and in practice -- many core practices such as informed consent,
confidentiality and anonymity that form the traditional building blocks
of ethical research (Conway, 2021; Markham and Buchanan, 2015; Zimmer,
2010, 2018). Digital research is not intrinsically more risky than
traditional face-to-face research. Rather it is how we understand risk
to participants, researchers, and institutions that have fundamentally
changed (Eynon et al., 2009). This can have implications for the design
and conduct of ethical research practices that the meet the shared
principles of respect, beneficence, and justice underpinning ethical
standards and requirements (Tiidenberg, 2017).

The online environment has blurred the lines between private and public,
creating significant ethical dilemmas for researchers. In Australia, the
\emph{Privacy Act 1988} regulates the fundamental human right of privacy
aiming to protect personal information, defined as `information or an
opinion about an identified individual or an individual who is
reasonably identifiable' (section 6) (The Law Library, 2018). In the
digital context, identifying what reasonably constitutes personal
information is rarely clear cut. Scholars have called into question the
norms associated with the ``very essence of privacy'' (Marwick and Boyd,
2014). danah boyd argued it is no longer possible to conceive of a
binary distinction between public and private in the online environment.
Data should instead be understood as part of a socially messy process
focused on control over the flow of information (`Privacy, Publicity,
and Visibility', 2010).

Concepts of power and control are further reflected in Helen
Nissenbaum's lens of ``contextual integrity''. From this perspective,
any approach must respect users' privacy by identifying and taking into
consideration the expected norms and roles associated with the flow of
information in any given context (Nissenbaum, 2004). Here privacy
becomes an ``ongoing negotiation of contexts in a networked ecosystem in
which contexts regularly blur and collapse'' (Marwick and Boyd, 2014:
1063). This is particularly true for social media data where studies
indicate wide discrepancies between users understandings and the
technological realities of data collection, storage, and sharing
(Fiesler and Proferes, 2018; Kemp, K., Gupta, C., Campbell, M., 2024;
Markham et al., 2018; Markham and Buchanan, 2015; Tiidenberg, 2017).
These different understandings, alongside the architectural
specificities of platforms optimised to encourage and entice users to
share information publicly, contribute to a privacy paradox. This may
reflect the dissonance between users' expressions of concern/care about
privacy and their actual online behaviours. With no clear guidance
available, and the layers of complexity involved in demarcating private
and public information, researchers are arguably compelled to explore
harms and risks on a case-by-case basis (Franzke et al., 2020).

Further complexities arise in relation to differing demarcations of what
constitutes public and private online data depending on the location of
a researcher. For example, across the European Union's (EU) member
states researchers are subject to the rules and definitions provided in
the General Data Protection Regulation (GDPR). The definition of
personal data adopted in this legislation is similar in the UK and in
Australia (the location of this paper's case study) where data that
contains ``information or an opinion about an individual's ``racial or
ethnic origin, political opinions or associations, religious or
philosophical beliefs, trade union membership or associations, sexual
orientation or practices, criminal record, health or genetic
information, or some aspects of biometric information'' (OAIC, 2023) are
classified as sensitive personal information under the \emph{Privacy Act
1988}. This is very different to the USA and Canada where online data is
frequently characterised under Common Law as public negating
requirements for ethical review. In other jurisdictions definitions can
be narrower. Because of the narrower definitions held in Australia, UK
and across the EU, researchers in these jurisdictions must provide
evidence as to how personal information will be protected. This can
include or necessitate applications for ethics approval from a REC/IRB.
At times the complex nature of the privacy issues have led RECs to adopt
a relatively risk-adverse approach impacting on the granting of
permission (Whelan, 2018).

Social media data also raises challenges for the process of informed
consent, which is a core part of human ethics processes used to protect
deontological norms of autonomy, equality, and respect. Informed consent
is predicated on users' ``autonomy, competence, and ability to
understand risk; and assumptions of it being possible for researchers to
imagine and predict future harm\ldots{} {[}such as{]}...storing data in
a cloud.'' (Tiidenberg, 2017: 470). There are circumstances where
informed consent is understood to be unfeasible and/or result in greater
risk to those engaged in the research process. In these cases, the
National Statement on the Ethical Conduct of Research (National
Statement) and REC may grant a waiver by addressing nine criteria in
section 2.3.10 of the National Statement (National Health and Medical
Research Council, the Australian Research Council and Universitities
Australia. Commonwealth of Australia, Canberra, 2007 (Updated 2018):
23).

However, applying these nine criteria to online data can be challenging.
Ensuring ``sufficient protection of their privacy'', for example,
necessitates engaging with complex issues around privacy and
de-identification techniques. It raises questions about what
``sufficient'' may mean for online users. One approach used to
demonstrate ``sufficient protection'' is to rely on a platform's Terms
of Service (ToS), used by digital service providers to set the
rules/guidance users must agree to prior to use of the online service.
ToS incorporate information pertaining to the collection, sharing, and
storage of data. Nevertheless, ToS have been shown to be deeply
problematic, especially when used as a proxy for informed consent. For
example, a study by Obar and Oeldorf-Hirsch sampled 543 online users and
found most individuals often ignored ToS policies on social networking
services, and even in cases where individuals did navigate to the ToS
policy page they remained on the page ``just long enough to scroll to
the `accept' button'' (Obar and Oeldorf-Hirsch, 2020: 140). Many people
seldom consider the long-term record made by their online actions and
posts, suggesting a contradiction between expectations and ToS
(Mackinnon, 2022). Data collection, even when done in compliance with
ToS, is contrary to the socially derived norms on a platform and the
privacy expectations of users (Daly et al., 2019). However, studies have
also shown how some social media companies use overly technical language
and volume of content as a means of obfuscating ToS, reducing the
likelihood of users reading and comprehending what information will be
collected and shared (Fiesler and Proferes, 2018; Franzke et al., 2020;
Kemp, K., Gupta, C., Campbell, M., 2024). Similarly, researchers have
highlighted problems with accepting as `neutral' the constraints set by
social media companies that govern access to users' data. These factors
raise questions around the transparency of data provided by social media
companies and power imbalances resulting from the financial disparities
involved in who can afford to purchase data (Bruns, 2019; Fiesler et
al., 2020; Kinder-Kurlanda et al., 2017).

Ethics requirements for confidentiality and anonymity are meant to
provide research participants with a sense of security, particularly
when sharing sensitive or personal information. Anonymisation of
research participants has become normalised through most ethical codes
of practice (Gerrard, 2021). In the online environment, however, it is
far harder to anonymise or robustly de-identify users. Studies have
demonstrated how easy it is to both reverse a de-identification process
and reconstruct the identity of a user through the combination of
different datasets (Barbaro et al., 2006; Narayanan and Shmatikov,
2008). Users may also be unaware of how the aggregation of data can be
used by researchers to generate insights into broader behavioural
traits. A person may consent to a tweet being `public', but be far less
happy if that tweet is used to identify broader tweeting patterns that
could be used to influence them in ways they had not anticipated
(Lauterwasser and Nedzhvetskaya, 2023). The de-identification process
may also risk compromising the integrity of data, rendering it less
useful for research purposes (Tiidenberg, 2017).

Another issue relating to privacy and confidentiality that is specific
to the digital environment arises from the use of pseudonymous
identities as a means of facilitating non-identifiable content. In some
cases, the platform or discussion board architecture require or
encourage the use of pseudonyms. In other cases, the practice is used
for reasons ranging from perceived safety or to facilitate the posting
of deceptive and hateful content (Gerrard, 2021). The use of pseudonyms,
particularly in the context of informed consent, raise ethical
quandaries for researchers. Attempts to uncover the real-world
identities behind a pseudonym could result in considerable harm to the
user and the researcher, especially where the practice is connected to
problematic content such as hate speech, violent extremism and
terrorism. Some scholars have suggested that where it is possible to
connect a pseudonym to a real-world user, they should be offered the
opportunity to maintain the use of the pseudonym during the proposed
research (Tiidenberg, 2017). Other researchers, such as Yip and
colleague, argued that the use of pseudonyms in the carding forums they
retrieved using a hacked dataset -- carders.cc, meant that it was not
possible for them to obtain informed consent (Thomas et al., 2017: 452).
Because of the considerable ethical challenges involved in seeking to
identify users adopting the practice of pseudonymous online engagement,
this is an area that arguably requires considerably more robust public
discussion among researchers utilising online data.

\section{HOW THESE RISKS ARE EXACERBATED WITH HACKED DATA}

The ethical challenges and ensuing risks examined above are further
problematised when examined in the context of the potential use of
hacked data in research. There are a growing number of hacked data that
have the potential to contribute to public interest. Journalists and
activists have been quick to utilise these datasets in a wide range of
investigations, but in the academic domain there has been considerable
less uptake of hacked datasets (Michael, 2015). In part, this likely
reflects ethical concerns about the illicit acquisition of the data.
Questions arise as to whether subsequent use of hacked data implicitly
condones or even encourages future acts of hacking (Tiidenberg, 2017).
Other scholars have noted that while hacked data remains unethical from
a moral perspective, there may well be times when exemptions can be
permitted to allow its ethical use for purposes of public interest
research (Ienca and Vayena, 2021a). The National Statement, however,
makes no reference to the use of hacked data, and the criteria it
establishes to obtain waivers of consent are hard to apply to these
cases. Tailored guidelines such as the Internet Research: Ethical
Guidelines 3.0 merely recognise the subject as necessitating further
consideration. Looking for insights from scholars who have published
research that draws from hacked data is equally underwhelming. In their
comprehensive review of studies using illicit data, Thomas and
colleagues note how most of the studies using hacked data provided
limited insights into the ethical processes, debates, and discussions
that informed their decisions (Thomas et al., 2017). This lack of formal
and informal guidance on the ethics of using hacked data creates a
problematic vacuum for researchers and RECs seeking best practices
around the specific risks and harms involved in the use of hacked data.

An alternative way to understand online privacy is as a process that
involves control and power over the flow of information. How the public
release of hacked data fundamentally changes the dynamics of this
process was of critical importance to our consideration of the use of
data stolen and released publicly from online platforms associated with
far-right violent extremists. Two events from 2015/16 demonstrate issues
relevant to researchers using stolen data: the hack of the controversial
infidelity site, Ashley Madison, and the scrapping of the more
conventional dating site, OkCupid. Despite the normative differences
between these sites and those of far-right extremist groups, these cases
highlight important considerations for researchers.

The Ashley Madison case stands as an example of the considerable harms
that can occur when sensitive data perceived by users as being private
becomes widely accessible to the public. Following the release of the
dataset, attempts to `out' users of the site for having extramarital
affairs and/or engaging in homosexual relations resulted in relationship
breakdowns, imprisonment in countries where homosexuality is banned, and
suicide (Cross et al., 2019; Mansfield-Devine, 2015). The dataset
remains online, yet there is little evidence academics have engaged with
the data itself. In the one example we found, the authors explicitly
noted the harms done previously to users of the site and noted they had
``handled and processed these data with the utmost concern for personal
security and privacy'' (Chohaney and Panozzo, 2018: 70). The lack of
controversy following the article's publication likely reflects the
active engagement of the authors in considering ethical issues and
applying deidentification processes to the data. It may also reflect the
temporal and behavioural distance of the authors from the act of the
hack itself (Egelman et al., 2012). This stands in contrast to the next
example where it was the academic researchers themselves who acquired
and published the data.

In 2016, two Danish graduate students published a dataset they had
scraped from the online dating site OKCupid. Although OKCupid is a less
controversial dating site than Ashley Madison, the publication of
publicly identifiable information including usernames, age, location and
insights into users' sexual preferences by two academics sparked media
controversy (Zimmer, 2016). When questioned about the data release, the
two students argued they had reproduced publicly available information.
In doing so, they appeared to conflate accessibility of data with an
inherent publicness. Aarhus University subsequently distanced itself
from the researchers, although no disciplinary and/or criminal action
was ever taken against them. Open Science Framework (OSF), who had
hosted the dataset, later removed it following a legal claim filed by
OKCupid under the Digital Millenium Copyright Act (DMCA) (Hackett,
2016). This legal action is noteworthy because OKCupid had previously
granted two researchers, Kim and Escobedo-Land, permission to access
data from the dating site that also led to the publication of
identifiable information (Kim and Escobedo-Land, 2015). The inclusion of
the data in an academic journal as opposed to a publicly available data
hosting platform such as the OSF likely contributed to the fact this
presence remained unchallenged for many years. In 2021, Xiao and Ma
wrote to the original journal explaining they had successfully asked the
original academics to redact the publicly identifiable data while
raising their concerns more broadly about the failure of the journal
(and academics) to rigorously engage with the ethical issues around
informed consent, perceptions of privacy, and control over the data
raised by the incident (Xiao and Ma, 2021).

For many researchers however, accessing robust digital data is becoming
increasingly difficult. This reflects the introduction of increasingly
defensive positions around data sharing by companies such as Meta
(Perriam et al., 2020). Similarly, actions by X (formally Twitter) that
implemented higher charges for researchers seeking limited access to its
Application Programming Interface (API)\footnote{An Application
  Programming Interface (API) are the rules and protocols that
  facilitate interaction with a software application. The API allows two
  different software systems to communicate with each other. Researchers
  seeking data from online platforms often rely on APIs as they can
  specify what data is received, the data is returned in a structured
  format, are relatively reliable and usually more compliant with a
  platform's terms of service.}, and Meta's decision to wind down its
social monitoring tool CrowdTangle that provided limited snapshot data
into public engagement across Facebook, are actions that impede
collection within these mainstream and popular social media platforms
(Berger, 2023; Bruns, 2019; Scire, 2024). Retrieving data from the
increasingly popular encrypted messaging apps such as Signal and
Telegram is also challenging for legal (ToS), ethical (what constitutes
private in these spaces), and technological (knowledge) reasons (Tuck et
al., 2023). In these contexts, hacked data can provide researchers with
public interest information otherwise difficult to access. Nevertheless,
public interest does not, in itself, mitigate other ethical obligations
and considerations.

Providing a record of how and from where data have been obtained is
critical for ethical research, including research utilising hacked data,
affording scholars' opportunities to assess the veracity of the
research. Transparency has been identified as important specifically in
the sub-discipline of online violent extremism and terrorism, given
prior critiques relating to the quality and substance of the empirical
data being used to generate theories and insights (Conway, 2017). Hacked
data poses significant challenges. For example, it is not always
possible to know who and how a hack was carried out. Identifying changes
to the data made by the hackers is possible, but identifying parts of
the data not released or hacked is harder. The illegal nature of hacking
means there are risks in publicly outing a hacker, if not made public
previously. In these circumstances, perhaps the most effective approach
is to detail, if public, who hacked the data and where it was obtained
from.

\section{CASE STUDY BACKGROUND}

The ethical dilemmas detailed in the following case study arose during
the PhD project of one of the authors. Building on existing scholarship,
the study examines the behaviours and language of far-right violent
extremists in the online environment to identify online signals of
extremist mobilisation to violence (Bradbury et al., 2017; Kernot et
al., 2022; Scrivens et al., 2023; Smith, 2021; Wojciechowski et al.,
2024). The research questions explore the potential identification of
online behaviours that may be positively associated with offline
mobilisation to violence. To do this, the project tests a hypothesis
developed by Brown and colleagues using the same datasets that
identified three online behaviours that indicated mobilisation to
violent action (Brown et al., 2024). As discussed further below, the
hacked datasets provide a relatively large cohesive body of information
created by a hard-to-reach community of users -- far-right violent
extremists. The Iron March and Gab hacked data sets used for the studies
have been shown to contain users who not only engage with the beliefs
and ideas associated with far-right violent extremism but have also
mobilised to offline violence. Accessing these datasets provides a
unique opportunity to explore potential behavioural differences between
processes of radicalisation to beliefs and ideas, and mobilisation to
violent action -- our central research interest.

The discussion board Iron March has been described as an ``online
incubator'' and the ``the primary organizational platform for a
transnational neo-fascist accelerationist terrorist network\ldots''
(Upchurch, 2021). It has also been linked to the formation of numerous
`real world' far-right violent extremist groups, including in Australia.
The site went offline in 2017. In 2019 the entire SQL database was
published online. Originally uploaded to the Internet Archive, the data
was later removed but remains widely accessible. Gab is a `fringe'
social media site with minimal moderation policies focused on `free
speech'. Gab has been connected to various far-right violent extremists
including Robert Bowers who posted his intent to carry out a violent
attack on the site prior to murdering eleven people in the Tree of Life
Synagogue, Pittsburgh USA, October 2018. In 2021 the CEO of Gab, Andrew
Torba, confirmed the platform had been breached. The hacked data is
currently held by Distributed Denial of Secrets (DDoS) who have
restricted the full release of the data to verified journalists and
researchers. We successfully applied for access to DDoS.

The use of our ethics process as a case study provides an opportunity to
deeply engage with the theoretical framework provided by Ienca and
Vayena for researchers seeking to explore the viability of balancing the
risks involved with the potential use of hacked data against the
benefits. Ienca and Vayena outline six criteria researchers should
consider when weighing up the possible ethical use of hacked data; (1)
risk-benefit, (2) uniqueness, (3) consent, (4) traceability, (5)
privacy, and (6) approval from an IRB (or in our case an REC) (Ienca and
Vayena, 2021a). These criteria map well against the core concepts of
most ethical guidelines and provided a useful starting point for our
ethical considerations about the use of the Iron March and Gab data. As
such, we believe these criteria could be usefully incorporated into
existing tailored ethical guidelines used by researchers as a starting
point for their own ethical considerations, and by REC/IRB as a
framework to consider the benefits and risks of hacked data ethical
applications. Reflecting the stance taken by Ienca and Vayena, we do not
see the sixth criteria -- approval from an ethics board as optional
(Ienca and Vayena, 2021b). Our location in Australia mandates approval
by an REC given the presence of sensitive personal information in the
hacked data, as defined by the \emph{Privacy Act 1988}.

Franzke et al. remind scholars that engaging in ethical decision making
requires not just asking the right questions but also developing
practices that encourage ``dialogical reflection'' (Franzke et al.,
2020: 6--7). To assist in these processes, we considered numerous
ethical frameworks developed by researchers engaged in the digital
environment before utilising the framework provided by Ienca and Vayena.
Zimmer notes that regardless of whether these are presented as
relatively proscriptive or more situationally reflexive, most guidelines
appear to coalesce around the three core ethical principles of respect,
benefice, and justice (Zimmer, 2018). It is therefore possible to see
these principles reflected in guidelines such as the Menlo report,
developed by the United States Department of Homeland Security and the
AoIR's IRE 3.0 as well as in the more reflexive and situational
approaches adopted in Gliniekca's situated ethics framework for Reddit
and Nissenbaum's nine-stage decision making heuristic (Franzke et al.,
2020; Gliniecka, 2023; Homeland Security Science and Technology, 2012;
Nissenbaum, 2009). These latter frameworks remind researchers of the
importance of recognising the human relationships that are at the heart
of the creation of digital data.

Importantly, none of these frameworks are presented by their authors as
proscriptive. Rather they seek to help guide and frame the
identification of the right ethical questions and associated ethical
discussions and debates that facilitate a weighing up of the risks and
benefits associated with the research. As we explored different ethical
frameworks, we noted that many incorporated a desirability for
incorporating participant engagement into the collection and analysis of
digital data. In the context of hacked data however, engaging with
individuals whose information has been stolen can be harmful for both
user and researcher for a variety of reasons explored in more depth in
the sections below. Our search of the literature identified Ienca and
Vayena's framework as uniquely designed to address the specific concerns
associated with hacked data (Ienca and Vayena, 2021a). While Boustead
and Herr, and Thomas et al. highlighted overarching issues relating to
the use of hacked data, such as issues of informed consent, transparency
and accountability and harms to vulnerable populations, neither provided
an overarching structured approach (Boustead and Herr, 2020; Thomas et
al., 2017). This was particularly important to us because our location
in Australia and the type of data incorporated in our hacked datasets
created a legislative requirement for University REC approval.

Although the framework presented by Ienca and Vayena is structured
around six topics, it also emphasized the importance of reflexive
situational discussion around these different issues. These dual
qualities helped us to explore the specific ethical challenges arising
from hacked data to better consider the risks and benefits associated
with the research. Ultimately the framework, like so many of those
developed by digital researchers, seeks not to generate a proscriptive
or utilitarian calculation of risks and benefits, but to facilitate
critical debate about how the risks and benefits relate to humans and
human relationships and how, if at all, potential harms from research
can be mitigated throughout the life cycle and beyond of a research
project. It is not our intention to suggest that this framework would be
appropriate for all researchers of hacked data or that other frameworks
are less valid. Rather, our aim was to address the ongoing request by
digital scholars to provide real-life case studies of complex ethical
issues relating to the digital environment so as to afford researchers
with reference points as to ways these issues could be addressed (Clark
et al., 2019: 69).

\section{BALANCING RISKS vs BENEFITS}

Perhaps the most foundational of the six topic areas set out by Ienca
and Vayena focuses on the balancing of risks and benefits arising from
the collection, use, and storage of hacked online data. In our case, the
most relevant issues revolved around balancing the risks related to the
theft of the data from the users, platforms, and the platform owners,
with the public benefit of advancing knowledge about a movement
supporting the use of violence to fundamentally undermine principles of
equality, the rights of women and minorities, and the rule of law.

Online far-right violent extremism and terrorism are perceived by
counter-terrorism scholars, law enforcement and intelligence
practitioners as playing a facilitative role in an ongoing threat to
democracy and the rule of law (ASIO, 2020; Conway, Scrivens and McNair,
2019). While there are many different groups that fall within the
umbrella term `far-right violent extremism', we define it as referencing
a pro-white racial/ethnic/sexual in-group identity that is hatefully
exclusionary to a perceived out-group that is non-white and often
focused on Jews, Muslim, immigrants, refugees, feminists, LGBTQIA+
(Berger, 2018; Scrivens et al., 2021). Those within this milieu are
proactively seeking the demise and at times violent destruction of
liberal democratic society. Increasing our understandings, as our
project seeks to do, into the how, why, and when questions that relate
to engagement with far-right violent extremism is critical if we are to
effectively prevent and manage the threats and consequences of violent
extremist activities.

The relationship between the online environment and engagement with
far-right violent extremism is known to be complex, but the empirical
body of evidence remains limited (Conway, 2017; Gill et al., 2017). Our
study aims to build on the emerging scholarship examining the
relationships between online language, emotions and attitudinal shifts
towards the acceptance of violence (Gill and Corner, 2015; Kernot et
al., 2022; Scrivens et al., 2021, 2023; Smith, 2018, 2021). Within the
study, the patterns identified in the analysis of the hacked data are
mapped against behavioural indicators in three structured professional
judgement tools used by practitioners when considering an individual's
potential threat to public safety by acting violently. There is a
well-documented lack of research into the efficacy of these tools
(Cubitt and Wolbers, 2022).

Scholars have raised concerns with research amplifying far-right violent
extremism to the benefit of the extremists (Ballsun-Stanton et al.,
2020; Phillips, 2018). As will be discussed below, our research
specifically seeks to minimise the potential of identifying any
individual associated with the hacked data. It focuses on structures and
patterns (rather than individuals). While this is not without its own
risks, there are significant benefits to carrying out public interest
research into far-right violent extremists given the harms involved in
engaging with these ideologies, groups, and networks online and offline.
The benefits threshold, we argue, can be achieved through stringent risk
mitigation processes, including those outlined in this paper.

\section{UNIQUENESS OF THE DATA}

Ienca and Vayena further point to the uniqueness of the data as a key
criterion to consider for ethical use of hacked data. In this vein, Poor
\& Davidson argued against the ethical use of hacked Patreon data noting
that ``we want this data but we don't need it'' (Poor and Davidson,
2016). In this specific case, the data was not unique enough to justify
its use given they were able to source similar but less ethically
compromised data (Poor and Davidson, 2016). However, given the aims of
our research project, there is a strong case to support the use of the
hacked Iron March and Gab data. To explore the ethical considerations,
we asked ourselves whether other sources of information could be
identified that would negate the need to use the hacked data. We then
considered the risks and benefits arising from the use of these
alternative sources and if they compared with those associated with the
hacked data.

The collection of systematic primary data from violent extremists and
terrorists has long been the Achillies heel of Terrorism and Violent
Extremism Studies (Sageman, 2014). The sample size of individuals
engaged with these ideologies, groups, and networks have thankfully been
relatively small. Some individuals are incarcerated which provides its
own challenges in terms of accessing data. The potential involvement of
these individuals in illicit activities aimed at the demise and/or
violent destruction of liberal democracy makes engagement potentially
harmful for both participants and researchers. The advent of the digital
environment and the vast amount of data it generates was seen as an
opportunity to collect relevant primary information from far-right
violent extremists who have a long history of using digital technologies
for communication, recruitment, and mobilisation to violence (Berlet,
2001; Conway, Scrivens and Macnair, 2019). However, as noted previously,
collecting online data has become more difficult over the years.

The fringe environment where far-right extremists are often active is
relatively volatile. Platforms can shut with no warning following
disruption by network hosting providers and network infrastructure
companies. Parler, for example, was created in 2018 as a free-speech
alternative social media platform. Its association with violent content
led to Amazon and others refusing to host the site and a series of
shutdowns. Users migrated to other free messaging sites (Williams et
al., 2021). Iron March is believed to have shut down following its hack
in 2017 and sites seeking to fill the gap left, such as Fascist Forge,
have since shut down. Using hacked data in our own project will
therefore limit the risks of the platform going offline, changing the
ToS, or users migrating in significant numbers during the collection
period.

Technological changes, social trends and digital affordances online have
fundamentally changed since Iron March was hacked in 2017 contributing
to the dataset's uniqueness. Iron March is an example of a web 1.0
message board, a type of platform that has largely decreased in use
since the arrival of web 2.0 social media platforms and messaging apps.
From a technical perspective a hack of a web 2.0 site such as Telegram
or Discord would more likely result in channel compromise, rather than
the all-of-service compromise that occurred at iron March. In addition,
despite the current relaxing of online moderation around hate speech,
operational security has become a far more salient issue for users
engaged in and with violent extremist content online. These behavioural
shifts also contribute to the unique nature of the dataset as users of
Iron March, unlike those on sites such as Fascist Forge that emerged
later, appeared to engage in communications without concerns around
observation and/or investigation by academics, journalists and law
enforcement and without concerns about moderation. The Iron March
dataset exists as a function of its community, the relative lack of
operational security, and the technologies of the time making it very
difficult to replicate a similar all-of-service dataset. It provides our
project unique access to a comparatively unfiltered dataset of online
behaviours relating to far-right violent extremist users, some of whom
have engaged in real-world violence.

The Gab hack provides another unique comprehensive snapshot into a
platform closely associated with far-right violent extremism. Accessing
data from Gab has become harder with the platform introducing measures
to actively restrict scraping of the site in 2023 with its chief of
technology announcing how ``through constant monitoring of the site and
the sources of traffic we are able to identify specific IPs and networks
where scraping is originating. We ban these sources, and continue to ban
new ones as they appear'' (\emph{Gab News}, 2023). The current ToS on
Gab raise legal concerns around any attempt to acquire data from the
platform above and beyond personal, non-commercial use with any other
type of use requiring requests to be sent to the platform's legal
services (Terms of Service - Gab Social, 2025). These restrictions make
the hacked Gab data a useful insight into a platform associated with
far-right violent extremism that also remains popular today (Collins,
2024). Another unique element of the Gab data is the inclusion of
public-facing and private posts. These allow us to explore an
under-developed area of research examining how different online spaces
(public/private) may impact on the behaviours, language, and emotions of
users in relation to attitudes to violence. Collectively, we concluded
that it was not possible to replicate the data through other sources. As
such, the hacked data from Iron March and Gab met the threshold of
`unique'.

\section{INFORMED CONSENT}

The acquisition of informed consent is another central component in
Ienca and Vayena's framework. In the context of hacked datasets, it is
critical to consider the potential risks and benefits that may arise
when seeking consent, not just in relation to the users, platforms, and
platform owners, but also in relation to the welfare of the researchers
and their institutions. It is also necessary to consider the
appropriateness and possibility of acquiring a waiver of consent from an
IRB/REC. In our case, we argued that acquiring informed consent from the
users in the hacked dataset would have been impractical and potentially
harmful.

The size and number of user accounts in the two hacked datasets
presented challenges to acquiring consent. The Gab data contained over
40 million posts relating to 1000s of different user accounts and, while
smaller, the Iron March dataset still contained over 1500 user accounts.
The hacked nature of the data also made seeking compliance with the
platforms ToS redundant. Even if the size of the datasets were smaller,
obtaining consent would have necessitated the researchers proactively
seeking to identify and/or re-identify all users in the dataset.

Furthermore, Gab and Iron March are closely associated with the
production of hateful, violent, and in some jurisdictions illegal
content. Identifying and/or re-identifying users could therefore create
considerable harm around user and/or researcher welfare. Some users may
not have been aware of the theft of their data and raising awareness
could inadvertently cause users' distress. For example, some users may
have subsequently moved away from far-right violent extremist narratives
and communities. Contacting and identifying these users could induce
psychological harms including the generation of negative emotions such
as shame and fear or even spark a re-engagement with past beliefs and
communities. On the other hand, some users may welcome or encourage
identification with far-right violent extremist platforms, meaning
identification could risk amplifying hateful content to novel audiences,
something the researchers are keen to avoid (Phillips, 2018). It is also
unlikely that users involved in these types of platforms would consent
to research that is (correctly) perceived to be detrimental to their
anti-democratic objectives (Fuchs, 2022). In this sense, the researchers
agreed to take a normative position regarding the value of liberal
democratic principles.

The presence of personal information in the datasets means that seeking
to re-identify users, even for purposes of informed consent, would place
the researchers in direct convention of the responsibilities set out in
the \emph{Privacy Act 1988}. So while Post reminds us that ``to use
{[}hacked{]} data without the consent of those who were violated is to
violate the violated anew'' (Post, 1991), in our case we believe seeking
consent would arguably do more harm. Seeking a waiver that mandates the
development of processes to protect the privacy and confidentiality of
the users involved in the hack, is arguably a more ethically justifiable
approach.

Finally seeking to engage with users who were active in spaces connected
to far-right violent extremism can create vulnerabilities for us as
researchers. Academics involved in the study of online violent extremism
have reported a range of harmful behaviours including online and offline
targeting and harassment as a consequence of their research (Conway,
2021; Frangou, 2019). In our case, our identities also reflect those
that are actively targeted and vilified by far-right extremists.
Utilising datasets that have already been collected and published
online, including hacked datasets, can help reduce some of these risks
by limiting researchers' exposure too and involvement with online
platforms and environments associated with violent extremists.

The use of such datasets by no means fully mitigate the risks associated
with doing research on far-right violent extremism, they provide an
alternative and arguably less risky avenue of engagement. Additional
safeguarding around the process of engagement with the data were also
put in place, including time limits for viewing the data, mental health
first aid assessment processes and trauma informed practice techniques.
Although Ienca and Vayena do not explicitly reference researcher
welfare, it is important to consider the potential risks that arise
particularly when engaging with complex and potentially risky
communities such as those within the far-right violent extremist
spectrum. These concerns to researchers, particularly those facing early
career researchers have been articulated in the research studies by
Vaughan and colleagues and Pearson and colleagues exploring issues
around researchers' security, safety and resilience (Pearson et al.,
2023; Vaughan et al., 2024). More broadly, the AoIR have also released
guidance for researchers using data from the increasingly complex online
environment (AoIR Risky Research Working Group, 2025). In our specific
case -- hacked data relating to anti-democratic, and potentially violent
illegal behaviours and activities -- we argued a waiver from consent
would be more effective in protecting privacy and confidentiality of
those who's personal information was contained in our datasets
\emph{and} the researchers involved in the proposed project(s). As a
result, we successfully applied for a waiver for consent from our REC.

PRIVACY

Ethical use of data requires researchers to protect the privacy and
confidentiality of participants and this requirement should be extended
to participants found in hacked data (Ienca and Vayena, 2021a). Our
ethical reflections considered questions about the manifestation of
relationships of power and control arising from the hacking and use of
hacked data. We asked what measures of privacy could be feasibly applied
to the data to minimise further impact on those involved in the hack. As
hacked data is often publicly accessible, we also considered what
measures of deidentification could be applied, what additional risks
these actions could result in, and what risks remained.

The hacking of Iron March and Gab breached the user's privacy and
confidentiality. If we use the data, we must therefore both acknowledge
this and be ethically comfortable with it, while also minimising further
harms. Under the \emph{Privacy Act 1988} the hacked datasets contain
information classified as sensitive personal information. Users on both
sites post content containing information about ethnicity, political
opinions, religious beliefs and affiliations, philosophical beliefs,
sexual practices and orientations. To protect the sensitive personal
information and considering our application for a waiver from consent,
we designed our research project to include a comprehensive data
management agreement covering storage, access to and subsequent
archiving/sharing of the data -- although the Gab data will not be
shared in-line with the ToS that guided our acquisition of the data from
DDoS. Whilst we acknowledge the imperfect science that is
de-identification, it still provides a degree of risk mitigation. Prior
to commencing our data analysis, we committed to carrying out a process
of de-identification to replace each unique identifiers/usernames with a
de-identified reference.

Adopting de-identification processes to protect the privacy of the users
allows us to recognise the layers of privacy -- real and perceived --
incorporated into the creation of the information in our datasets. Iron
March users, for example, have been described as having perceived their
engagement on the site to be far more private than on subsequent sites
such as Fascist Forge (Scrivens et al., 2023). This sense of being
unobserved may have encouraged users to post more hateful and extreme
content (Upchurch, 2021). The social norms of Gab encourage users to
adopt pseudonyms when posting content to the site. Research has
indicated that perceptions of anonymity online can contribute to a sense
of disinhibition that results online in more toxic and aggressive
posting behaviours and activities (Nitschinsk et al., 2023). Users may
or may not regret posting this type of content over time, they may have
genuine concerns about being identified in the real world with far-right
violent extremists -- de-identification helps to mitigate these
potential harms whilst protecting the privacy and confidentiality of the
users.

The public nature of hacked data has meant research has already been
conducted on both datasets resulting in the publication of identifying
details that relate to some of the users of these sites (Owen and Hume,
2019; \emph{Security Magazine}, 2021). Furthermore, given the
accessibility of the hacked data online there are risks around the
robustness of any de-identification processes adopted to protect the
privacy of users. A mitigating factor is the lens of focus for our
research. Our work explores behavioural and language patterns across the
datasets as opposed to identifying individual specificities. Findings
associated with the research will reflect this lens of inquiry
reinforcing the risk mitigation of the de-identification processes. In
addition, presentations and publications resulting from our research
will avoid the use of quoted content, where this cannot be avoided the
quotes will be redacted or paraphrased to restrict opportunities for
re-identification.

Iron March and Gab have both been associated with illegal activity.
There are risks our research could identify illegal behaviours connected
to violent extremist activity. The historical nature of the data and its
existence in the public domain for some time limit the chances of the
team identifying novel illicit behaviours and/or activities. The
de-identification processes adopted further mitigates the risk of our
ability to pass on personally related information to law enforcement.
Should specific illegal activities be associated with a potential threat
to public safety be identified, these will be passed on to law
enforcement as is ethically appropriate and required.

TRACEABILITY

Finally, Ienca and Vayena (2021a) point to the need for researchers to
be transparent regarding the traceability of hacked data, including how
and where all data has been obtained. In this sense, the framework
understands and presents traceability as a linear mechanism ``that
allows individuals and organizations to forensically `follow' data along
a life cycle\ldots''(Thylstrup, 2022: 664) for purposes of transparency.
Nevertheless, it is important to recognise this linear reading of
traceability that perceives data sets as `neutral sites of knowledge
retrieval' (Thylstrup, 2022: 658), has been problematised and critiqued
by scholars, seeking a more culturally complex, iterative and expanded
understanding of the data set.

These arguments for a more complex reading of traceability reflect our
own experience when considering the ethical use of the Iron March and
Gab data sets. While we followed a procedural approach, documenting both
our own distance from the initial theft of the data and the subsequent
steps taken to acquire it, we felt this was by itself insufficient to
meet an ethical threshold. Arguably, the illicit nature of the data
demands greater ethical reflexivity that goes beyond a simple emphasis
on documenting provenance. Grappling with the complex power dynamics and
cultural aspects of our research (Mackinnon, 2022; Ogden, 2022),
especially given the hacked data, we continually revisited the ethical
arguments for studying far-right extremists. We also reflected on our
own position as researchers, committed to liberal values like equality
and human rights. Because these ethical considerations are captured in
the other criteria presented by Ienca and Vayena, it provided a
framework for iterative reflexive practices, as opposed to just process,
that helped reinforce our own ethical positionality. However,
traceability remains an area where further research focused on its
implications in the context of hacked data would be of benefit. While
many of these issues are arguably captured in the other criteria
presented in the framework, traceability remains an area where further
research focused on its implications in the context of hacked data would
be of benefit.

CONCLUSION: The case for (reflexive) ethical review

Hacked data represents an ethically grey area for researchers. The act
of stealing data is illegal, violating users' privacy and
confidentiality and potentially creating significant harms. Despite the
innate illegality of hacked datasets, there are, as this paper explores,
a range of potential benefits that can arise from the use of hacked
datasets, particularly in the context of research focused on difficult
to access violent extremist communities that seek to disregard the human
rights of others. Understanding how the potential benefits balance
against the numerous risks associated with hacked data demand engagement
with the reflexive and dialogical ethical approach (Franzke et al.,
2020: 4). An important part of our success in achieving ethics approval
related to the ongoing support we received from the university's REC
Chair who took time prior to submission to advise on our concerns and
mitigations. We encourage robust and open pre-emptive discussions with
ethics committees given the potential unfamiliarity of committee members
with the challenges arising from the use of digital data and hacked data
for research purposes. However, as the researchers engaged in this paper
discovered, practical guidance on the development of a reflexive ethical
practice to guide the balancing of risks and benefits associated with
hacked data remains limited. This situation raises challenges given the
inherent moral issues associated with data that has been obtained
illegally and without consent.

This paper has provided a detailed case study outlining the pragmatic
challenges involved when applying the six theoretical criteria proposed
by Ienca and Vayena in their exploration of circumstances that may allow
for the ethical use of hacked data. The use of a single case study
provided us with the opportunity to deeply engage with the theoretical
ethical framework proposed by Ienca and Vayena as a mechanism to
facilitate the balancing of risks vs benefits. Providing insight into
the reflexive process we engaged in responds to a call within the
research community for researchers to use case studies to make visible
the lived experience of negotiating through ethical conundrums in
research. In this way we hope that our paper contributes to academic
discussions around the philosophy of research ethics, including those
that consider how competing frameworks across different country's
IRBs/RECs may apply to researchers seeking to use hacked data.

The paper also provides an alternative perspective to the case study
presented by Poor and Davidson, who found against the ethical use of a
proposed hack dataset in their research. In this way, the paper adds to
the well-documented requirement for discussion of the ethical practices
adopted in real world case studies in the context of the complex issues
raised when using online data including hacked online datasets. Our case
study highlights how the overarching criteria proposed by Ienca and
Vayena can usefully structure the ethical discussions that arise when
researchers consider the ethical use of hacked data. By including the
key issues that framed our own thinking when grappling with the risks
and potential mitigation strategies, we believe the paper offers a
helpful insights for others who may also be considering the use of
hacked data. In our case, we felt the arguments for using the data and
the risk mitigation strategies we could apply, were stronger than those
against not using the data. We acknowledge risks do remain. Challenges
involved in de-identification processes creates an ongoing risk users
will be `outed' and associated with far-right violent extremism in ways
that are harmful to their psychological and physical well-being. In
addition, when we publish on our research, we risk being negatively
targeted by far-right violent extremists. Despite these risks, we
believe these are balanced against the overall benefits of the research
to society and are commensurate with ethical obligations for researchers
to push back against movements that seek to undermine, or indeed
completely dismantle, the human rights of others.

The ethical practices we have developed to mitigate the risks afford us
the opportunity to present this as a case study where hacked data can be
used ethically. We began this article highlighting the grey areas of
ethical conduct and the sense of unease researchers face when using
online data and in particular hacked data. These feelings have not been
silenced and nor, we argue, should they be. Instead, we need to amplify
the discomforts that arise when using hacked data -- it is only through
these types of debates and discussions that reflexive ethical practices
are developed, critiqued, and developed some more.

\section*{REFERENCES}
\setlength{\parindent}{-0.2in}
\small
\setlength{\leftskip}{0.2in}
\setlength{\parskip}{8pt}
\vspace*{-0.1in}
\hspace{-0.2in}AoIR Risky Research Working Group (2025) An AoIR Guide to Researcher
Protection and Safety. Available at:
https://aoir.org/wp-content/uploads/2025/05/AoIR-Risky-Research-Report\_2025.pdf.

ASIO (2020) Director-General's Annual Threat Assessment \textbar{}
Australian Security Intelligence Organisation. Available at:
https://www.asio.gov.au/director-generals-annual-threat-assessment.html
(accessed 27 June 2020).

Ballsun-Stanton B, Waldek L, Droogan J, et al. (2020) \emph{Mapping
Networks and Narratives of Online Right-Wing Extremists in New South
Wales}. Available at: https://zenodo.org/record/4071472.

Barbaro M, Zeller T and Hansell S (2006) A face is exposed for AOL
searcher no. 4417749. \emph{New York Times} 9(2008): 8.

Berger JM (2018) \emph{Extremism}. Essential Knowledge Series. MIT
Press. Available at: https://mitpress.mit.edu/books/extremism (accessed
5 May 2020).

Berger JM (2023) \emph{The Last Twitter Census - Examining Baseline
Metrics for Twitter Users during a period of Dramatic Change}. Vox-Pol
Centre of Excellence. Available at:
https://voxpol.eu/wp-content/uploads/2024/01/DCUP10774-Vox-Pol-Report23-Digital-231212.pdf.

Berlet C (2001) When Hate went Online. In: \emph{Northeast Sociological
Association, Spring Conference}, 2001. Available at:
https://citeseerx.ist.psu.edu/document?repid=rep1\&type=pdf\&doi=6a2ecf050969ec4635068fa4154ba650f16c12f0.

Boustead AE and Herr T (2020) Analyzing the Ethical Implications of
Research Using Leaked Data. \emph{PS, political science \& politics}
53(3). Cambridge University Press: 505--509.

Bradbury R, Bossomaier T and Kernot D (2017) Predicting the emergence of
self-radicalisation through social media: a complex systems approach.
\emph{Terrorists' Use of the Internet: Assessment and Response, IOS
Press, Amsterdam}: 379--389.

Brown O, Smith LGE, Davidson BI, et al. (2024) Online signals of
extremist mobilization. \emph{Personality \& social psychology
bulletin}. SAGE Publications: 1461672241266866.

Bruns A (2019) After the `APIcalypse': social media platforms and their
fight against critical scholarly research. \emph{Information,
Communication and Society} 22(11). Routledge: 1544--1566.

Chohaney ML and Panozzo KA (2018) Infidelity and the Internet: The
Geography of Ashley Madison Usership in the United States.
\emph{Geographical review} 108(1). Routledge: 69--91.

Clark K, Duckham M, Guillemin M, et al. (2019) Advancing the ethical use
of digital data in human research: challenges and strategies to promote
ethical practice. \emph{Ethics and information technology} 21(1).
Springer Science and Business Media LLC: 59--73.

Collins J (2024) The Gift of Gab: A Netnographic Examination of the
Community Building Mechanisms in Far-Right Online Space. \emph{Terrorism
and Political Violence}. Routledge: 1--20.

Conway M (2017) Determining the Role of the Internet in Violent
Extremism and Terrorism: Six Suggestions for Progressing Research.
\emph{Studies in Conflict and Terrorism} 40(1). Routledge: 77--98.

Conway M (2021) Online Extremism and Terrorism Research Ethics:
Researcher Safety, Informed Consent, and the Need for Tailored
Guidelines. \emph{Terrorism and Political Violence} 33(2). Routledge:
367--380.

Conway M, Scrivens R and McNair L (2019) Right-wing extremists'
persistent online presence: history and contemporary trends.
International Centre for Counter-terrorism-\/-The Hague. Epub ahead of
print 2019. DOI: 10.19165/2019.3.12.

Conway M, Scrivens R and Macnair L (2019) Right-wing extremists'
persistent online presence: History and contemporary trends.
\emph{Policy brief} . Epub ahead of print 2019. DOI: 10.19165/2019.3.12.

Cross C, Parker M and Sansom D (2019) Media discourses surrounding
`non-ideal' victims: The case of the Ashley Madison data breach.
\emph{International Review of Victimology} 25(1). SAGE Publications Ltd:
53--69.

Cubitt T and Wolbers H (2022) \emph{Review of violent extremism risk
assessment tools in Division 104 control orders and Division 105A
post-sentence orders}. Australian Institute of Criminology. Available
at: https://www.aic.gov.au/sites/default/files/2023-05/sr14.pdf.

Daly A, Kate Devitt S and Mann M (2019) \emph{Good Data}. Institute of
Network Cultures. Available at:
https://play.google.com/store/books/details?id=FKzAwwEACAAJ.

Egelman S, Bonneau J, Chiasson S, et al. (2012) It's Not Stealing If You
Need It: A Panel on the Ethics of Performing Research Using Public Data
of Illicit Origin. In: \emph{Financial Cryptography and Data Security},
2012, pp. 124--132. Springer Berlin Heidelberg. Available at:
http://dx.doi.org/10.1007/978-3-642-34638-5\_11.

Eynon R, Schroeder R and Fry J (2009) New techniques in online research:
challenges for research ethics. \emph{Twenty-first century society}
4(2). Informa UK Limited: 187--199.

Fiesler C and Proferes N (2018) ``Participant'' Perceptions of Twitter
Research Ethics. \emph{Social Media + Society} 4(1). SAGE Publications
Ltd: 2056305118763366.

Fiesler C, Beard N and Keegan BC (2020) No robots, spiders, or scrapers:
Legal and ethical regulation of data collection methods in social media
Terms of Service. \emph{Proceedings of the International AAAI Conference
on Web and Social Media} 14. Association for the Advancement of
Artificial Intelligence (AAAI): 187--196.

Flyvbjerg B (2006) Five misunderstandings about case-study research.
\emph{Qualitative inquiry: QI} 12(2). SAGE Publications: 219--245.

Frangou C (2019) The growing problem of online harassment in academe
---. Available at:
https://universityaffairs.ca/features/feature-article/the-growing-problem-of-online-harassment-in-academe/
(accessed 8 May 2024).

Franzke AS, Bechmann A, Zimmer M, et al. (2020) \emph{Internet Research:
Ethical Guidelines 3.0}. Association of Internet Researchers. Available
at: https://aoir.org/reports/ethics3.pdf.

Fuchs C (2022) `DEAR MR. NEO-NAZI, CAN YOU PLEASE GIVE ME YOUR INFORMED
CONSENT SO THAT I CAN QUOTE YOUR FASCIST TWEET?' Questions of social
media research ethics in online ideology critique. In: Fuchs C (ed.)
\emph{Digitial Fascism}. Routledge, pp. 385--394. Available at:
http://fuchs.uti.at/wp-content/socialmediaresearchethics.pdf.

\emph{Gab News} (2023) How Gab Deals With Scraping on the Open Web.
Available at:
https://news.gab.com/2023/07/scraping-the-open-web-gab-coms-approach/
(accessed 4 November 2023).

Gerrard Y (2021) What's in a (pseudo)name? Ethical conundrums for the
principles of anonymisation in social media research. \emph{Qualitative
research: QR} 21(5). SAGE Publications: 686--702.

Gill P and Corner E (2015) Lone actor terrorist use of the Internet and
behavioural correlates. In: Lee Jarvis, Stuart Macdonald, and Thomas M.
Chen (ed.) \emph{Terrorism Online: Politics, Law, Technology and
Unconventional Violence}. London UK: Routledge, pp. 35--53. Available
at:
https://www.taylorfrancis.com/chapters/edit/10.4324/9781315848822-10/lone-actor-terrorist-use-internet-behavioural-correlates-paul-gill-emily-corner.

Gill P, Corner E, Conway M, et al. (2017) Terrorist Use of the Internet
by the Numbers: Quantifying Behaviors, Patterns, and Processes.
\emph{Criminology \& public policy} 16(1): 99--117.

Gliniecka M (2023) The Ethics of Publicly Available Data Research: A
Situated Ethics Framework for Reddit. \emph{Social Media + Society}
9(3). SAGE Publications Ltd: 20563051231192020.

Hackett R (2016) Researchers Caused an Uproar By Publishing Data From
70,000 OkCupid Users. Available at:
https://fortune.com/2016/05/18/okcupid-data-research/ (accessed 21
February 2025).

Homeland Security Science and Technology (2012) \emph{The Menlo Report
Ethical Principles Guiding Information and Communication Technology
Research}. 3 August. Homeland Security Science and Technology .
Available at:
https://www.dhs.gov/sites/default/files/publications/CSD-MenloPrinciplesCORE-20120803\_1.pdf.

Ienca M and Vayena E (2021a) Ethical requirements for responsible
research with hacked data. \emph{Nature Machine Intelligence} 3(9).
Nature Publishing Group: 744--748.

Ienca M and Vayena E (2021b) Is it ethical to use hacked data in
scientific research? Available at:
https://papers.ssrn.com/abstract=3843733 (accessed 11 October 2023).

Kemp, K., Gupta, C., Campbell, M. (2024) \emph{Singled Out - Consumer
understanding - and misunderstanding - of data broking, data privacy,
and what it means for them}. February. Consumer Policy Research Centre
and UNSW . Available at:
https://cprc.org.au/wp-content/uploads/2024/02/CPRC-Singled-Out-Final-Feb-2024.pdf.

Kernot D, Leslie S and Wood M (2022) A community resilience linguistic
framework for risk assessment: using second order moral foundations and
emotion on social media. \emph{Journal of Policing, Intelligence and
Counter Terrorism}. Routledge: 1--18.

Kim AY and Escobedo-Land A (2015) OkCupid data for introductory
statistics and data science courses. \emph{Journal of statistics
education: an international journal on the teaching and learning of
statistics} 23(2). Informa UK Limited.

Kinder-Kurlanda K, Weller K, Zenk-Möltgen W, et al. (2017) Archiving
information from geotagged tweets to promote reproducibility and
comparability in social media research. \emph{Big Data \& Society} 4(2).
SAGE Publications Ltd: 2053951717736336.

Lauterwasser S and Nedzhvetskaya N (2023) Privacy in Public?: The Ethics
of Academic Research with Publicly Available Social Media Data.
\emph{Berkeley Jounral of Sociology}. Epub ahead of print 8 November
2023.

Mackinnon K (2022) Critical care for the early web: ethical digital
methods for archived youth data. \emph{Journal of Information
Communication and Ethics in Society} 20(3). Emerald: 349--361.

Mansfield-Devine S (2015) The Ashley Madison affair. \emph{Network
Security} 2015(9): 8--16.

Markham AN and Buchanan EA (2015) Internet Research: Ethical Concerns.
In: Wright JD (ed.) \emph{International Encyclopedia of the Social \&
Behavioral Sciences (Second Edition)}. Oxford: Elsevier, pp. 606--613.
Available at:
https://www.sciencedirect.com/science/article/pii/B978008097086811027X.

Markham AN, Tiidenberg K and Herman A (2018) Ethics as Methods: Doing
ethics in the Era of Big Data Research---introduction. \emph{Social
media + society} 4(3). SAGE Publications: 205630511878450.

Marwick AE and Boyd D (2014) Networked privacy: How teenagers negotiate
context in social media. \emph{New media \& society} 16(7). SAGE
Publications: 1051--1067.

Michael GJ (2015) Who's afraid ofWikiLeaks? Missed opportunities in
political science research: Who's afraid of WikiLeaks? \emph{The review
of policy research} 32(2). Wiley: 175--199.

Narayanan A and Shmatikov V (2008) Robust De-anonymization of Large
Sparse Datasets. Available at:
https://ieeexplore.ieee.org/document/4531148 (accessed 19 August 2024).

National Health and Medical Research Council, the Australian Research
Council and Universitities Australia. Commonwealth of Australia,
Canberra (2007 (Updated 2018)) \emph{National statement on Ethical
Conduct in Human Research 2007 (Updated 2018)}.

Nissenbaum H (2004) Privacy as Contextual Integrity. \emph{Washington
Law Review} 79(1): 119.

Nissenbaum H (2009) \emph{Privacy in Context: Technology, Policy, and
the Integrity of Social Life}. Stanford Law Books. Available at:
https://www.degruyterbrill.com/document/doi/10.1515/9780804772891/html
(accessed 30 May 2025).

Nitschinsk L, Tobin SJ, Varley D, et al. (2023) Why Do People Sometimes
Wear an Anonymous Mask? Motivations for Seeking Anonymity Online.
\emph{Personality \& social psychology bulletin}: 1461672231210465.

OAIC (2023) What is privacy? Available at:
https://www.oaic.gov.au/privacy/your-privacy-rights/your-personal-information/what-is-privacy
(accessed 15 August 2024).

Obar JA and Oeldorf-Hirsch A (2020) The biggest lie on the Internet:
ignoring the privacy policies and terms of service policies of social
networking services. \emph{Information, communication and society}
23(1). Informa UK Limited: 128--147.

Ogden J (2022) ``Everything on the internet can be saved'': Archive
Team, Tumblr and the cultural significance of web archiving.
\emph{Internet histories} 6(1--2). Informa UK Limited: 113--132.

Owen T and Hume T (2019) EXCLUSIVE: A U.S. Marine Used the Neo-Nazi Site
Iron March to Recruit for a `Racial Holy War'. Available at:
https://www.vice.com/en/article/exclusive-a-us-marine-used-the-neo-nazi-site-iron-march-to-recruit-for-a-race-war/
(accessed 27 August 2024).

Pearson E, Whittaker J, Baaken T, et al. (2023) \emph{Online Extremism
and Terrorism Researchers' Security, Safety, and Resilience: Findings
from the Field}. VOX-Pol. Available at:
https://doras.dcu.ie/28332/1/Online-Extremism-and-Terrorism-Researchers-Security-Safety-Resilience.pdf
(accessed 28 February 2025).

Perriam J, Birkbak A and Freeman A (2020) Digital methods in a post-API
environment. \emph{International journal of social research methodology}
23(3). Informa UK Limited: 277--290.

Phillips W (2018) \emph{The Oxygen of Amplification - Better Practices
for Reporting on Extremists, Antagonists, and Manipulators Online}. May.
Data \& Society . Available at:
https://datasociety.net/library/oxygen-of-amplification/ (accessed 20
February 2024).

Poor N and Davidson R (2016) The ethics of using hacked data: Patreon's
data hack and academic data standards. Available at:
http://www.datascienceassn.org/sites/default/files/Ethics\%20of\%20Using\%20Hacked\%20Data\%20-\%20Patreon\%E2\%80\%99s\%20Data\%20Hack\%20and\%20Academic\%20Data\%20Standards.pdf
(accessed 24 July 2023).

Post SG (1991) The echo of Nuremberg: Nazi data and ethics.
\emph{Journal of medical ethics} 17(1): 42--44.

`Privacy, Publicity, and Visibility' (2010). Available at:
https://www.danah.org/papers/talks/2010/TechFest2010.html (accessed 25
April 2024).

Sageman M (2014) The Stagnation in Terrorism Research. \emph{Terrorism
and Political Violence} 26(4). Routledge: 565--580.

Scire S (2024) A window into Facebook closes as Meta sets a date to shut
down CrowdTangle. Available at:
https://www.niemanlab.org/2024/03/a-window-into-facebook-closes-as-meta-sets-a-date-to-shut-down-crowdtangle/
(accessed 25 April 2024).

Scrivens R, Chermak SM, Freilich JD, et al. (2021) Detecting Extremists
Online: Examining Online Posting Behaviors of Violent and Non-Violent
Right-Wing Extremists. Available at:
http://dx.doi.org/10.37805/pn2021.21.remve.

Scrivens R, Osuna AI, Chermak SM, et al. (2023) Examining Online
Indicators of Extremism in Violent Right-Wing Extremist Forums.
\emph{Studies in Conflict and Terrorism} 46(11). Routledge: 2149--2173.

\emph{Security Magazine} (2021) Far-right platform Gab confirms it was
hacked. Available at:
https://www.securitymagazine.com/articles/94733-gab-confirms-it-was-hacked
(accessed 27 August 2024).

Smith D (2018) So how do you feel about that? Talking with Provos about
emotion. \emph{Studies in Conflict and Terrorism} 41(6). Taylor \&
Francis: 433--449.

Smith D (2021) Passionate Belief: Ideology, Emotion and Terrorist
Action. \emph{Emotions: History, Culture, Society} 5(1). Brill: 6--24.

Terms of Service - Gab Social (2025). Available at:
https://gab.com/about/tos (accessed 28 February 2025).

The Law Library (2018) \emph{Privacy Act 1988 (Australia) (2018
Edition)}. CreateSpace Independent Publishing Platform. Available at:
https://play.google.com/store/books/details?id=ORY7uAEACAAJ.

Thomas DR, Pastrana S, Hutchings A, et al. (2017) Ethical issues in
research using datasets of illicit origin. In: \emph{Proceedings of the
2017 Internet Measurement Conference}, New York, NY, USA, 1 November
2017, pp. 445--462. IMC '17. Association for Computing Machinery.
Available at: https://doi.org/10.1145/3131365.3131389 (accessed 10 April
2023).

Thylstrup NB (2022) The ethics and politics of data sets in the age of
machine learning: deleting traces and encountering remains. \emph{Media,
culture, and society} 44(4). SAGE Publications: 655--671.

Tiidenberg K (2017) Ethics in digital research. In: (Ed) FU (ed.)
\emph{The SAGE Handbook of Qualitative Data Collection} . London,
England: SAGE Publications, pp. 466--481.

Tuck H, Guhl J, Smirnova J, et al. (2023) \emph{Researching the Evolving
Online Ecosystem: Telegram, Discord and Odysee}. Institute for Strategic
Dialogue (ISD) and CASM technology. Available at:
https://www.isdglobal.org/wp-content/uploads/2023/04/Researching-the-Evolving-Online-Ecosystem\_Telegram-Discord-Odysee.pdf.

Upchurch HE (2021) The Iron March Forum and the Evolution of the ``Skull
Mask'' Neo-Fascist Network. \emph{CTC Sentinel}: 1--19.

Vaughan A, Braune J, Tinsley M, et al. (2024) The ethics of researching
the far right: Critical approaches and reflections. Vaughan A, Braune J,
Tinsley M, et al. (eds). Manchester University Press. Available at:
https://doi.org/10.7765/9781526173898 (accessed 28 February 2025).

Whelan A (2018) Ethics Are Admin: Australian Human Research Ethics
Review Forms as (Un)Ethical Actors. \emph{Social Media + Society} 4(2).
SAGE Publications Ltd: 2056305118768815.

White RE and Cooper K (2022) Case Study Research. In: \emph{Qualitative
Research in the Post-Modern Era}. Cham: Springer International
Publishing, pp. 233--285. Available at:
https://link.springer.com/content/pdf/10.1007/978-3-030-85124-8\_7?pdf=chapter\%20toc.

Williams HJ, Evans AT, Ryan J, et al. (2021) \emph{The Online Extremism
Ecosystem: Its Evolution and a Framework for Separating Extreme from
Mainstream}. December. RAND.

Wojciechowski TW, Scrivens R, Freilich JD, et al. (2024) Testing a
probabilistic model of desistance from online posting in a right-wing
extremist forum: distinguishing between violent and non-violent users.
\emph{International journal of comparative and applied criminal
justice}. Routledge: 1--20.

Xiao T and Ma Y (2021) A letter to the \emph{journal of statistics and
data science education} --- A call for review of ``OkCupid data for
introductory statistics and data science courses'' by Albert Y. kim and
Adriana Escobedo-land. \emph{Journal of statistics and data science
education: an official journal of the of the American Statistical
Association} 29(2). Informa UK Limited: 214--215.

Zimmer M (2010) ``But the data is already public'': on the ethics of
research in Facebook. \emph{Ethics and information technology} 12(4):
313--325.

Zimmer M (2016) OkCupid Study Reveals the Perils of Big-Data Science.
Available at:
https://www.wired.com/2016/05/okcupid-study-reveals-perils-big-data-science/
(accessed 21 February 2025).

Zimmer M (2018) Addressing Conceptual Gaps in Big Data Research Ethics:
An Application of Contextual Integrity. \emph{Social Media + Society}
4(2). SAGE Publications Ltd: 2056305118768300.

Zook M, Barocas S, Boyd D, et al. (2017) Ten simple rules for
responsible big data research. \emph{PLoS computational biology} 13(3):
e1005399.

\end{document}